\documentclass[11pt,twoside]{article}  
\usepackage{apn3conf} 
 
\begin{document}   

%
%
%
%
 
\title{Radio Cores of Bipolar Nebulae: Search for Collimated Winds} 
 
%
%
%
 
\author{Ting-Hui Lee} 
\affil{Department of Physics and Astronomy, University of Calgary, Canada} 
\author{Jeremy Lim, Sun Kwok\altaffilmark{1}} 
\affil{Institute of Astronomy and Astrophysics, Academic Sinica, Taiwan}  
\altaffiltext{1}{Also Department of Physics and Astronomy, University of Calgary, 
Canada} 
 
%
%
 
\contact{Ting-Hui Lee} 
\email{thlee@iras.ucalgary.ca} 
 
%
%
%
%
%
 
\paindex{Lee, T.-H.} 
\aindex{Lim, J.} 
\aindex{Kwok, S.} 
 
%
%
 
\authormark{Lee, Lim, \& Kwok} 
 
%
%
 
 
 
\begin{abstract}          
 
We present results of our search for collimated ionized winds in
bipolar nebulae using the Very Large Array (VLA) and the Australia
Telescope Compact Array (ATCA).  Our search is motivated by the
discovery of an ionized jet in the bipolar nebula M 2-9 (Lim \& Kwok
2003) that may be responsible for sculpting the nebula's
mirror-symmetric structure. To determine if such jets are a common
feature of bipolar nebulae, we searched for optically-thick radio
cores - a characteristic signature of ionized jets - in 11 northern
nebulae with the VLA at 1.3 cm and 0.7 cm, and in 5 southern nebulae
with the ATCA at 6 cm and 3.6 cm. Two northern objects, 19W32 and M
1-91, and two southern objects, He2-84 and and Mz 3, exhibit a compact
radio core with a rising spectrum consistent with an ionized jet.  Th
2-B exhibits a steeply falling spectrum characteristic of nonthermal
radio emission.  Here we present a preliminary analysis of these five
radio cores and discuss the implications of our results.
 
\end{abstract} 
 
%
%
 
\section{Introduction} 
 
    Bipolar nebulae are defined as axially symmetric planetary nebulae
    having two lobes with an `equatorial' waist (Schwarz, Corradi, \&
    Stanghellini 1992).  In the context of the``generalized wind-blown
    bubble,'' a fast and spherically-symmetric wind from the central
    post-AGB star sweeps up a slowly-expanding axially-symmetric
    envelope previously expelled by the progenitor red giant star to
    produce a bipolar nebula.  Such a model, however, can only produce
    bipolar nebulae with wide waists, not those with narrow
    ``pinched'' waists (Soker \& Rappoport 2000, and references
    therein). Soker \& Rappaport argued that the creation of bipolar
    nebulae with narrow waists requires a collimated wind, which in
    their model originates from a white dwarf companion accreting the
    wind of its red giant primary, i.e., a symbiotic star system. Such
    a collimated wind in the form of an ionized jet has indeed been
    discovered in the prototype narrow-waist bipolar nebula M 2-9 (Lim
    \& Kwok 2000, 2003). The radio core of M 2-9 has a spectral index
    of 0.67, which is in nearly perfect agreement with the spectral
    index of 0.6 expected from an isothermal outflow expanding at a
    constant velocity and opening angle (Reynolds 1986). To determine
    if central ionized jets are a common feature of bipolar nebulae,
    we observed a total of 16 objects listed as bipolar nebulae with
    very narrow waists in Table 1 of Soker \& Rappaport (2000).  Our
    sample comprises 11 northern objects, IRAS 07131-0147, M 1-16, NGC
    2818, NGC 6302, 19W32, HB 5, NGC 6537, M 3-28, M 1-91 M 2-48, and
    NGC 7026, observed with the VLA at 1.3 cm and 0.7 cm, and 5
    southern objects, He 2-25, He 2-36, He 2-84, Th 2-B, and Mz 3,
    observed with the ATCA at 6 cm and 3.6 cm.
 
\section{Results and Analysis} 
 
\subsection{VLA Survey} 
 
    We made the VLA observations in March 2002 when the telescope was
    in its A configuration.  To correct for rapid tropospheric phase
    variations, we used fast switching with a calibration cycle time
    of just 2--3 minutes. The data were reduced and analyzed using the
    AIPS package developed by the National Radio Astronomy
    Observatory.  We achieved an angular resolution of up to
    $\sim$0.08$''$ at 1.3 cm and $\sim$0.04$''$ at 0.7 cm.

    We found that NGC 6302, Hubble 5, and NGC 6537 exhibit extended
    radio emission that could not be properly mapped due to the lack
    of short baselines in our observations.  By contrast, 19W32 and
    M1-91 show a compact central source only at both 1.3 cm and 0.7
    cm. No emission was detected from the remaining six objects.  Note
    that this is the first time the centers of all these nebulae have
    been examined at such high angular resolutions at radio
    wavelengths.
 
    To determine the flux densities of the radio cores observed in
    19W32 and M1-91, we fit a two-dimensional Gaussian structure to
    the measurements by applying the task IMFIT to the clean map and
    OMFIT to the visibility data.  We fit the entire dataset at 1.3
    and 0.7~cm, and also a more restricted dataset at 0.7~cm chosen to
    have the same angular resolution as at 1.3~cm to check for a
    greater inclusion of any extended nebular emission in the larger
    synthesized beam.  We found that both procedures gave the same
    results within measurement uncertainties.  Thus, in Table~1, we
    list the results obtained by fitting the datasets over their
    entire uv range.
 
\begin{table} 
\vspace{-2mm} 
\begin{center} 
\label{tb:vla-flux} 
\caption{Flux densities for compact cores detected with VLA} 
\begin{tabular}{c|c|c c c} \hline \hline 
    Object & Fitting Process & 1.3 cm (mJy) & 0.7 cm (mJy) & Spectral 
    Index \\ \hline 
    19W32 & \begin{tabular}{c} 
            Visibility fit \\ Map fit 
	    \end{tabular} & 
            \begin{tabular}{c} 
            3.58 $\pm$ 0.18 \\ 3.48 $\pm$ 0.17 
	    \end{tabular} & 
	    \begin{tabular}{c} 
	    6.18 $\pm$ 0.69 \\ 4.76 $\pm$ 0.57  
	    \end{tabular} & 
	    \begin{tabular}{c} 
            0.81 $\pm$ 0.18 \\ 0.47 $\pm$ 0.19  
	    \end{tabular} \\ \hline 
    M 1-91 &\begin{tabular}{c} 
            Visibility fit \\ Map fit 
	    \end{tabular} & 
            \begin{tabular}{c} 
            2.97 $\pm$ 0.16 \\ 2.65 $\pm$ 0.15 
	    \end{tabular} & 
	    \begin{tabular}{c} 
	    5.11 $\pm$ 0.39 \\ 3.97 $\pm$ 0.38  
	    \end{tabular} & 
	    \begin{tabular}{c} 
            0.79 $\pm$ 0.14 \\ 0.64 $\pm$ 0.17  
	    \end{tabular} \\ \hline \hline 
\end{tabular} 
\end{center} 
\vspace{-4mm} 
\end{table} 
 
    The rising spectrum of both 19W32 and M1-91 suggest that their
    radio cores are produced by optically-thick free-free emission.
    Fits to the measured visibilities show that these cores are
    resolved along one dimension.  If we assume a circularly uniform
    source of the measured dimensions, we deduce brightness
    temperatures as listed in Table~2 that are a factor of a few below
    that expected for ionized winds (in the case of M2-9, T$_{\rm B}
    \geq$ 4000K).  On this basis, we anticipate that the radio cores
    are noncircular, suggesting a collimated wind or jet.  We plan
    observations at higher angular resolutions to properly map the
    structure of these radio cores.
 
\begin{table} 
\vspace{-2mm} 
\begin{center} 
\label{tb:vla-tmp} 
\caption{Deduced brightness temperatures for 19W32 and M 1-91} 
\begin{tabular}{c|c|c c c} \hline \hline 
    Object & Wavelength & Resolved Axis & Position Angle
    & T$_{B}$ \\ \hline 
    19W32 & \begin{tabular}{c} 
            1.3 cm \\ 0.7 cm 
	    \end{tabular} & 
            \begin{tabular}{c} 
            63 $\pm$ 11 mas \\ 51 $\pm$ 10 mas 
	    \end{tabular} & 
	    \begin{tabular}{c} 
	    51$^{\circ}~\pm$ 19$^{\circ}$ \\ 56$^{\circ}~\pm$ 46$^{\circ}$ 
	    \end{tabular} & 
	    \begin{tabular}{c} 
            2150 $\pm$ 540 K \\ 1650 $\pm$ 490 K
	    \end{tabular} \\ \hline  
    M 1-91 &\begin{tabular}{c} 
            1.3 cm \\ 0.7 cm 
	    \end{tabular} & 
            \begin{tabular}{c} 
            77 $\pm$ ~7 mas \\ 34 $\pm$ ~6 mas
	    \end{tabular} & 
	    \begin{tabular}{c} 
	    86$^{\circ}~\pm$ ~6$^{\circ}$ \\ 58$^{\circ}~\pm$ 12$^{\circ}$
	    \end{tabular} & 
	    \begin{tabular}{c} 
            1190 $\pm$ 170 K \\ 3120 $\pm$ 810 K 
	    \end{tabular} \\ \hline \hline 
\end{tabular} 
\end{center} 
\vspace{-4mm} 
\end{table} 
 
\subsection{ATCA Survey} 
 
   We conducted the ATCA observations in March 2003 with the telescope
   in its 6-km configuration. The data were reduced using the MIRIAD
   reduction package.  We attained an angular resolution of
   $\sim$$3''$ at 6 cm and $\sim$$1.5''$ at 3.6 cm.

   We found He 2-36 to exhibit extended emission only, whereas He 2-84
   and Th 2-B exhibit only a compact central source.  Mz 3 show both
   compact and extended emission.  No emission was detected from He
   2-25.  For He 2-84 and Th 2-B, we applied the task IMFIT to the
   clean map and UVFIT to the visibility data to determine the flux
   densities of their radio cores. In both cases, we assumed their
   radio cores to have a two-dimensional Gaussian structure. For Mz 3,
   the flux density was obtained only for the bright compact core in
   the clean map, as can be seen in Figure~1.  The results are listed
   in Table~3.
 
\begin{table} 
\vspace{-2mm} 
\begin{center} 
\label{tb:atca} 
\caption{Flux densities for compact cores detected with ATCA} 
\begin{tabular}{c|c|c c c} \hline \hline 
    Object & Fitting Process & 6 cm (mJy) & 3.6 cm (mJy) & Spectral 
    Index \\ \hline 
    He 2-84 & \begin{tabular}{c} 
            Visibility fit \\ Map fit 
	    \end{tabular} & 
            \begin{tabular}{c} 
            12.1 $\pm$ 0.16 \\ \ 8.6 $\pm$ 0.06 
	    \end{tabular} & 
	    \begin{tabular}{c} 
	    \ 9.2 $\pm$ 0.16 \\ \ 7.7 $\pm$ 0.08  
	    \end{tabular} & 
	    \begin{tabular}{c} 
            $-$0.47 $\pm$ 0.04 \\ $-$0.18 $\pm$ 0.02  
	    \end{tabular} \\ \hline 
    Th 2-B &\begin{tabular}{c} 
            Visibility fit \\ Map fit 
	    \end{tabular} & 
            \begin{tabular}{c} 
            14.9 $\pm$ 0.12 \\ 15.1 $\pm$ 0.09 
	    \end{tabular} & 
	    \begin{tabular}{c} 
	    \ 8.2 $\pm$ 0.17 \\ 10.1 $\pm$ 0.06  
	    \end{tabular} & 
	    \begin{tabular}{c} 
            $-$1.01 $\pm$ 0.04 \\ $-$0.68 $\pm$ 0.02  
	    \end{tabular} \\ \hline 
     Mz 3  & Map pixel sum & 15.3 $\pm$ 0.14 & 18.3 $\pm$ 0.15 & $+$0.35 
            $\pm$ 0.02 \\ \hline \hline 
\end{tabular} 
\end{center} 
\vspace{-4mm} 
\end{table} 
 
\begin{figure}[ht] 
\plotfiddle{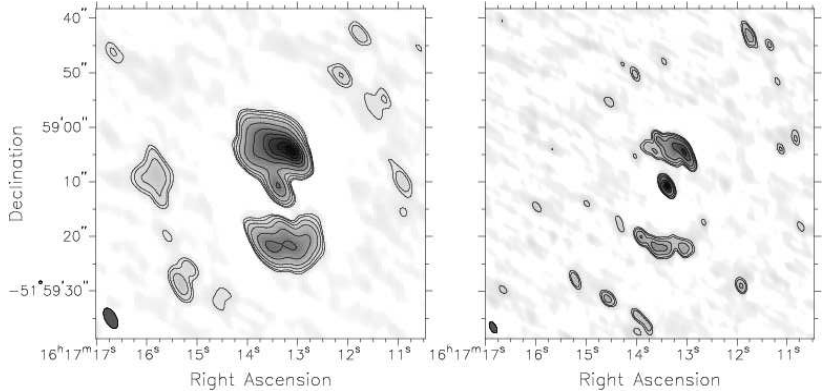}{50mm}{270}{50}{50}{-160}{145} 
\caption{Radio map of Mz 3 at 6 cm (left) and 3.6 cm (right). The 
  contours start at 3$\sigma$.} 
\end{figure} 
 
    Two of the three radio cores have spectral indices significantly
    smaller than the value of $-0.1$ expected for optically thin
    free-free emission.  However, these indices assume the flux
    measurements are of the cores only.  If diffuse emission is
    present on scales the array can detect, it will contaminate these
    fluxes, perhaps even in different amounts at 3.6 and 6 cm.  To
    guard against this possibility, we excluded short baselines that
    may be sensitive to extended emission, and also the long baselines
    at 3.6 cm to provide the same angular resolution as at 6~cm.  The
    results obtained from the clean maps are listed in Table~4.
 
\begin{table} 
\vspace{-2mm} 
\begin{center} 
\label{tb:atca-uv} 
\caption{Compact ATCA core fluxes vs. uv-range} 
\begin{tabular}{c|c|c c c} \hline \hline 
    Object & uv-range (k$\lambda$) & 6 cm (mJy) & 3.6 cm (mJy) & 
    Spectral Index \\ \hline 
    He 2-84 & \begin{tabular}{c} 
            50-100 \\ 65-100 \\ 75-100 
	    \end{tabular} & 
            \begin{tabular}{c} 
            1.83 $\pm$ 0.14 \\ 1.39 $\pm$ 0.16 \\ 1.08 $\pm$ 0.18 
	    \end{tabular} & 
	    \begin{tabular}{c} 
	    2.60 $\pm$ 0.10 \\ 1.89 $\pm$ 0.12 \\ 1.42 $\pm$ 0.15  
	    \end{tabular} & 
	    \begin{tabular}{c} 
            $+$0.60 $\pm$ 0.15 \\ $+$0.52 $\pm$ 0.22 \\ $+$0.47 $\pm$ 0.34  
	    \end{tabular} \\ \hline 
    Th 2-B &\begin{tabular}{c} 
            65-100 \\ 75-100 
	    \end{tabular} & 
            \begin{tabular}{c} 
            11.1 $\pm$ 0.16 \\ 10.8 $\pm$ 0.18 
	    \end{tabular} & 
	    \begin{tabular}{c} 
	    \ 5.7 $\pm$ 0.13 \\ \ 5.5 $\pm$ 0.16  
	    \end{tabular} & 
	    \begin{tabular}{c} 
            $-$1.13 $\pm$ 0.05 \\ $-$1.15 $\pm$ 0.06  
	    \end{tabular} \\ \hline 
     Mz 3  & \begin{tabular}{c} 
            45-100 \\ 55-100 
	    \end{tabular} & 
            \begin{tabular}{c} 
            13.3 $\pm$ 0.11 \\ 10.8 $\pm$ 0.14 
	    \end{tabular} & 
	    \begin{tabular}{c} 
	    19.5 $\pm$ 0.11 \\ 20.3 $\pm$ 0.11  
	    \end{tabular} & 
	    \begin{tabular}{c} 
            $+$0.65 $\pm$ 0.02 \\ $+$1.07 $\pm$ 0.02  
	    \end{tabular} \\ \hline \hline  
\end{tabular} 
\end{center} 
\vspace{-4mm} 
\end{table} 
  
    The difference in the core flux densities and resultant spectral
    indices evaluated using the different methods suggest that the
    emission comes a diffuse source and a compact source in the cases
    of He2-84 and Mz 3.  In these two cases, the more careful analysis
    incorporating the same uv-coverage (Table 4) indicates that we
    have detected an optically-thick radio core.  For Th 2-B, in both
    analyses we derive a spectral index close to $-1$, implying a
    nonthermal emission mechanism.  We plan observations at higher
    angular resolutions, as is possible with the newly-commissioned
    12~mm and 3~mm systems at the ATCA, to better understand the
    nature of these radio cores.

\acknowledgments 
 
    This work is supported by a grant from Natural Sciences and 
    Engineering Research Council of Canada.  
  
%
%
%
%

 
\end{document}